\documentclass[12pt]{article}

\usepackage{latexsym,amssymb,euscript}
\usepackage[dvips]{graphicx}
\usepackage{amsmath}
\usepackage{graphbox}
\usepackage{amsfonts}
\usepackage{amssymb, epsfig} 

\usepackage{float}
\usepackage{array}

\usepackage[dvipsnames]{xcolor}

\usepackage[utf8]{inputenc}

\usepackage{color,soul}

\begin{document}

\begin{center}
{\Large \textbf{Confinement-Higgs and deconfinement-Higgs transitions in
four-dimensional $SU(2)$ LGT at finite temperature}}

\vspace*{0.6cm}
\textbf{B. All\'es\footnote{email: alles@pi.infn.it}} \\
\vspace*{0.1cm}
\centerline{\it INFN Sezione di Pisa, Largo Pontecorvo 3, 56127 Pisa, Italy}
\vspace*{0.3cm}
\textbf{O.~Borisenko\footnote{email: oleg@bitp.kyiv.ua}} \\
\vspace*{0.1cm}
\centerline{\it INFN Gruppo Collegato di Cosenza, Arcavacata di Rende, 87036 Cosenza, Italy}
\centerline{\rm and}
\centerline{\it N.N.Bogolyubov Institute for Theoretical Physics,}
\centerline{\it National Academy of Sciences of Ukraine, 03143 Kyiv, Ukraine}
\vspace*{0.3cm}
\textbf{A. Papa\footnote{email: papa@fis.unical.it}} \\
\vspace*{0.1cm}
\centerline{\it Dipartimento di Fisica, Universit\`a della Calabria}
\centerline{\rm and}
\centerline{\it INFN Gruppo Collegato di Cosenza, Arcavacata di Rende, 87036 Cosenza, Italy}
\end{center}

%\vspace*{1cm}

\begin{abstract}
  We re-examine by numerical simulation the phase structure of the $(3+1)$-dimensional $SU(2)$
lattice gauge theory (LGT) with gauge fields coupled to Higgs fields at finite temperature.
  Concretely, we explore two different order parameters which are able to distinguish the three
phases of the theory: (i) the Fredenhagen-Marcu operator used to discriminate between deconfinement and confinement/Higgs phases and (ii) the Greensite-Matsuyama overlap operator proposed recently to distinguish confinement and Higgs phases. 
\end{abstract}

\section{Introduction}

In this paper we shall study the $(3+1)$-dimensional $SU(2)$ gauge-Higgs theory regularized on the lattice. The dynamical variables of this model  are: (i) one Higgs field at every site $x$, denoted $V(x)$, and (ii) one gauge field on each link $l$, denoted $U(l)\equiv U_n(x)$ if $l=(x,n)$ is the link departing from $x$ in the direction $n$. Both gauge and Higgs fields are taken in the fundamental representation. The partition function of the model on $\Lambda\in Z^{(d+1)}$ 
is defined as 
\begin{equation}
\label{gauge_higgs_pf}
Z_{\Lambda}(\beta, \gamma) \equiv Z 
=  \int \prod_l dU(l) \ \int \prod_x dV(x) \ e^{S_G + S_H} \;,
\end{equation} 
with the gauge $S_G$ and the Higgs $S_H$ actions given by 
\begin{eqnarray}
\label{gauge_action} 
S_G &=& \frac{1}{2} \ \beta \sum_{x,n<m} \ {\rm Tr} \ 
U_n(x) U_m(x+e_n) U^{\dagger}_n(x+e_m) U^{\dagger}_m(x)  \ ,   \\ 
\label{higgs_action}   
S_H &=& \frac{1}{2} \ \gamma \sum_{x,n} \ {\rm Tr} \
V^{\dagger}(x) U_n(x) V(x+e_n) \ . 
\end{eqnarray} 
Here, $x=(x_1,x_2,x_3,t)$ with $t\in [0,N_t-1]$, and $x_i\in [0,L-1]$ for $i=1,2,3$ denotes a site on the lattice $\Lambda$, while $e_n$ is the unit vector in the direction $n$.

Periodic boundary conditions are imposed in all directions.
The model is invariant under the action of local $SU_{\rm l}(2)$ transformations $\omega(x)$ and global  $SU_{\rm gl}(2)$ transformations $\alpha$,
\begin{eqnarray}
\label{zn_transform}
U_n(x) &\rightarrow& U_n^{\prime}(x) = 
\omega(x) U_n(x) \omega^{\dagger}(x+e_n) \ ,  \\ 
V(x) &\rightarrow& V^{\prime}(x) = \omega(x) V(x) \alpha  \ .
\end{eqnarray}
Thus, the full symmetry group is $SU_{\rm l}(2)\times SU_{\rm gl}(2)$. 

This model has been extensively studied at zero temperature and its phase diagram is well known (for a recent review of the physics of many gauge-Higgs LGTs see Ref.~\cite{pisa_rev_25}). It exhibits two phases, the confinement and the Higgs phases, separated by a thermodynamic transition in a restricted region of the phase diagram, where the gauge coupling $\beta$ is sufficiently large and the Higgs coupling $\gamma$ is sufficiently small. This first order transition and related second order endpoints have been analyzed in some detail in~\cite{fm_su2_zero_temp}. 
Since in other regions of the phase diagram the free energy and its derivatives are analytic functions of model parameters, there are no thermodynamic phase transitions in such regions~\cite{seiler_78,fradkin_shenker}.

Physics differs in the two phases. In the confinement phase one observes string tension and flux tube formation at short and intermediate distances, while at large distances a string breaking phenomenon takes place. There are no massless gauge excitations: gauge fields form bound states like in QCD. 
In the Higgs phase the strings and flux tubes do not appear because the gauge fields become massive due to a gauge-invariant Higgs mechanism~\cite{gauge_inv_higgs}. 
Thus, since no charged states appear in any of the two phases, in some sense both phases can be associated with confinement. The conventional confinement phase can be viewed as a charge-separation confinement phase (also called S-confinement), and the Higgs phase as a color confinement phase (C-confinement). The notion of the S-confinement was proposed and elaborated in Ref.~\cite{greensite_17_Sconf}. 

Due to the different physics and the absence of any thermodynamic transition between the S-confinement and C-confinement regions, it seems natural to ask whether an order parameter able to distinguish between the two phases could exist. 
It turns out that it is indeed possible to construct such a parameter by resorting to the  analogy between the gauge-Higgs action (\ref{higgs_action}) and Hamiltonians for spin-glass models~\cite{edwards_75}. Under this viewpoint, the gauge fields in (\ref{higgs_action}) play the role of
position-dependent coupling constant with a probability distribution weighted by the Wilson action. This analogy has been deeply explored in
Refs.~\cite{greensite_18_overlap,greensite_20_overlap} for the $SU(2)$ gauge-Higgs theory and a gauge-invariant analog of the Edwards-Anderson overlap operator for spin-glasses has been developed in these papers. Monte Carlo computations of that
operator did reveal a nontrivial critical behavior which is attributed to the spontaneous symmetry breaking of the global $Z(2)$ subgroup acting only on the Higgs fields (the so-called custodial symmetry). Thus, a quantity able to distinguish confinement from Higgs phases exists and the
Higgs phase is realized as a glassy phase with a nonzero value of the overlap operator, while the confinement phase is viewed as a paramagnetic phase with vanishing overlap operator. For more details we refer the reader to reviews~\cite{greensite_22_overlap,greensite_25_rev}. In spin-glass models the overlap operator of~\cite{edwards_75} distinguishes ferromagnetic phases from glassy phases. In gauge-Higgs theories the overlap operator of \cite{greensite_18_overlap,greensite_20_overlap} should be able to distinguish the deconfinement phase from the Higgs phase. This is indeed the case as it was shown in~\cite{Ward_22} for the five-dimensional $SU(2)$ gauge-Higgs theory and in~\cite{z2_gauge_Higgs} for the three-dimensional $Z(2)$ gauge-Higgs theories. 

\begin{figure}[htb]
\centering
\includegraphics[width=0.6\linewidth,clip]{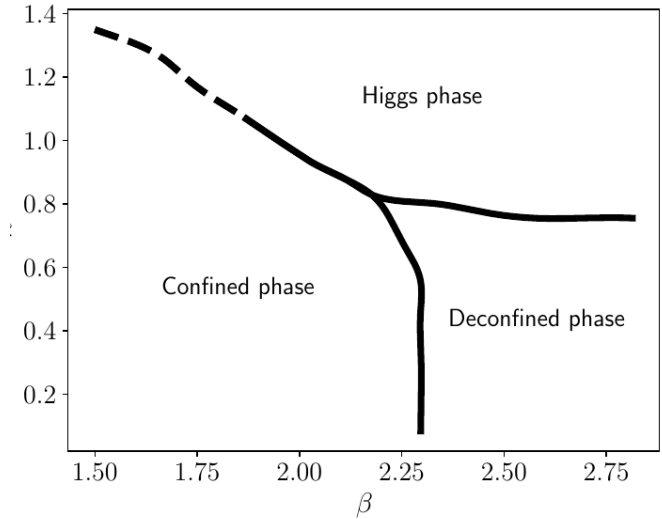}
\caption{Phase diagram of the $SU(2)$ gauge-Higgs theory on the lattice at finite temperature in the $(\beta,\gamma)$-plane for $N_t=4$. See text for details.}
\label{fig:phase_diagram_general}
\end{figure}

The phase diagram of the $SU(2)$ gauge-Higgs model exhibits a richer structure at finite temperature as shown in Fig.\ref{fig:phase_diagram_general} for $N_t=4$~\cite{su2_higgs_fin_temp}\footnote{This plot is taken from Ref.~\cite{su2_higgs_fin_temp} with permission of the authors.}. Specifically, a new deconfined phase appears at large $\beta$ and small $\gamma$. In the pure gauge theory, where $\gamma$ vanishes,
the critical coupling equals $\beta_c\approx 2.299$ for $N_t=4$ and $\beta_c\approx 2.51$ for $N_t=8$~\cite{pure_su2_fin_temp}. The lines separating the deconfined phase from the confinement and Higgs phases were determined numerically in~\cite{su2_higgs_fin_temp} for $N_t=4$. It remains unclear, however, if these lines represent a thermodynamic transition or just a rapid crossover. Much like in the theory at zero temperature, the transition weakens for decreasing $\beta$ and eventually turns into a crossover between confinement and Higgs phases. This crossover appears in Fig.\ref{fig:phase_diagram_general} as a dashed line.

There exists a non-local order parameter which can distinguish deconfinement from confinement and Higgs phases, at least in theories at zero temperature. This is the Fredenhagen-Marcu (FM) operator introduced in Ref.~\cite{FM_86}. This operator tests charged states in vacuum. In the confinement/Higgs phases the operator tends to a constant which is a function of $\beta$ and $\gamma$, while in the deconfined phase it decays to zero as the size $R$ of the operator increases (the precise definitions are given in Sec.~\ref{observables}). 
The rigorous proof supporting this behavior of the FM operator was constructed for the four-dimensional $U(1)$ model in~\cite{fm_4d_u1} and for the $Z(2)$ model in~\cite{FM_24}. Numerical simulations have been performed for $Z(2)$ gauge-Higgs theory in~\cite{z2_gauge_Higgs,FM_86_MC} and for $SU(2)$ in~\cite{fm_su2_zero_temp}. All simulations support  the expected behavior of the FM operator. These analytic and numerical results have been obtained for zero-temperature theories.

In this paper we study the behavior of the overlap and the FM operators in the $SU(2)$ gauge-Higgs model (\ref{gauge_higgs_pf}) at finite temperature. Our main goal is to reveal if these operators can suffice to discriminate between the three different phases of the model. We are not going to delve into details regarding the critical behavior as we only wish to understand the efficacy of these operators as non-local order parameters at finite temperature. 

The paper is organized as follows.  In Sec.~2 we define the overlap and the FM  operators involved in the analysis. In this section a simple non-rigorous argument is also outlined to justify why the FM operator may vanish in the deconfined phase of the theory. In Sec.~3 the results of our simulations are presented and Sec.~4 summarizes our conclusions.

\section{Order parameters}
\label{observables}

\subsection{Overlap operator}
\label{overlap_sec}

For testing purposes we have utilized two different definitions of the overlap operator. The first one, which we will denote by $G_{1s}$, was introduced in~\cite{greensite_20_overlap}, and consists of performing $n_{\rm sym}$ {\it special} Monte Carlo updatings on a copy of the running configuration
in equilibrium at the assigned values of $\beta,\gamma$. The word ``special'' has a twofold meaning: first, the updating must act on all gauge and Higgs fields except for the spatial links lying on a given slice, {\it e.g.}, $t=0$, at a fixed value of the temporal coordinate. Second, those updatings must be slow, that is, the successive configurations must not be too decorrelated. This is due to the fact that the overlap measures a transient phenomenon. Thus, if a heavy decorrelation were applied, like a cluster algorithm (where available) or by repeating many hits of more conventional local algorithms like Metropolis or Heat Bath, then one would end up with a vanishing $\langle G_{1s}\rangle$ at any value of the couplings $\beta,\gamma$.

Once the above $n_{\rm sym}$ special Monte Carlo updatings have been applied, the operator to be evaluated is
\begin{equation}
  G_{1s}\equiv\frac{1}{V_{\rm space}}\sum_{\vec{x}}\left(\frac{1}{n_{\rm sym}}\sqrt{\frac{1}{2}{\rm Tr}\left[\Phi(0,\vec{x})^\dagger\Phi(0,\vec{x})\right]}\right) \ , \ 
  \Phi(0,\vec{x})\equiv\sum_{{m}=1}^{n_{\rm sym}}V^{(m)}(0,\vec{x})\;,
  \label{overlapoperator}
\end{equation}
where $V^{(m)}(0,\vec{x})$ is the result of measuring the Higgs field 
at spatial position ${\vec{x}}$ over the $t=0$ slice after $m$ special Monte Carlo updatings,  with $1\leq m\leq n_{\rm sym}$, and $V_{\rm space}=L^3$ is the spatial volume of the slice.

The second definition, which will be denoted $G_w$, is similar to the first one but the gauge fields are fixed over the whole lattice during all data-taking special Monte Carlo sweeps. The overlap operator becomes 
in this case
\begin{equation}
	G_{w}\equiv\frac{1}{V_{\rm space}N_t}\sum_{\vec{x},t}\left(\frac{1}{n_{\rm sym}}\sqrt{\frac{1}{2}{\rm Tr}\left[\Phi(t,\vec{x})^\dagger\Phi(t,\vec{x})\right]}\right) \ , \ 
	\Phi(t,\vec{x})\equiv\sum_{{m}=1}^{n_{\rm sym}}V^{(m)}(t,\vec{x})\;,
	\label{overlapoperator2}
\end{equation}
where again $V^{(m)}(t,\vec{x})$ is the result of measuring the Higgs field at the spatial position $\vec{x}$ and time $t$ after $m$ hits of special Monte Carlo updatings, with $1\leq m\leq n_{\rm sym}$. Like for $G_{1s}$, also for $G_w$ the special Monte Carlo steps must be slow.

In principle, $n_{\rm sym}$ can be tuned in order to choose the best estimator. But a simpler strategy consists of repeating the measurements of (\ref{overlapoperator}) and
(\ref{overlapoperator2}) for a chosen set of
values of $n_{\rm sym}$ and fitting the results to the functional form
\begin{equation}
  G_a=\langle  G_a \rangle + \frac{A_a}{\sqrt{n_{\rm sym}}}\;,
  \label{overlapoperator_fit}
\end{equation}
where $a$ labels the two operators (\ref{overlapoperator}) and (\ref{overlapoperator2}), that is $a=1s$ or $w$, and
$\langle{G}_a\rangle$ indicates the physically meaningful result, while $A_a$ are constants. The functional form (\ref{overlapoperator_fit}) is expected from general statistical arguments. 
We also expect that $\langle G_{1s} \rangle$ and $\langle G_{w} \rangle$, in the large volume and $n_{\rm sym}\to\infty$ limits, lead to the same value for the critical couplings.

It is worth noting that each special Monte Carlo step for the measurement of $G_w$ is 
approximately four times faster than the measurement of $G_{1s}$. This is because the evaluation of $G_w$ is performed without the updatings of gauge fields.

%As stated above, a further important aspect of the measurement of (\ref{overlapoperator}) and
%(\ref{overlapoperator2}) is that the Monte Carlo updatings must be slow. This is due to the fact
%that the overlap operator measures a transient phenomenon. Thus, if a heavy decorrelation were applied (for instance by using a cluster algorithm or by repeating many
%hits of more conventional local algorithms like Metropolis or Heat Bath), one would obtain $\langle{G}_a\rangle$ zero for both $a$ and any value of the couplings $\beta,\gamma$.

\subsection{FM operator}
\label{2.2FM}

The FM operator is defined as follows, see~\cite{FM_86}. 
Let ${\cal{C}}$ be a closed rectangular loop of links with one side of length $T$
along the temporal coordinate and the other side of length $R$ along one spatial direction. We denote by $W({\cal{C}})$ the corresponding Wilson loop,
\begin{equation}
\label{wilson_loop_def}
W({\cal{C}}) \equiv  {\rm Tr} \ \prod_{l\in \cal{C}} \ U(l) \ . 
\end{equation}
By cutting $W({\cal C})$ into two halves along the temporal sides, we obtain a $\sqcap$ and a $\sqcup$ shaped chains of links. Let us take one of them, for instance
$\sqcap$, and call it ${\cal L}_{xy}$ where $x,y$ are the endpoints. Hence, the horizontal side of $\sqcap$ runs along a spatial direction and has length $R$, while the vertical sides run along the temporal direction and have lengths $T/2$. We then consider a gauge-invariant Wilson line, denoted by $V({\cal L}_{xy})$, and obtained by multiplying ${\cal L}_{xy}$ with the Higgs fields at the endpoints $x,y$,
\begin{equation}
\label{wilson_line_def}
V({\cal{L}}_{xy}) \equiv  {\rm Tr} \  V^{\dagger}(x) \ 
\Big( \prod_{l\in {\cal{L}}_{xy}} \ U(l) \Big) \ V(y) \ . 
\end{equation}
Given the $R$-dependent ratio
\begin{equation}
\label{FMoperator_def}
H(R,T) = 
\frac{\langle V({\cal{L}}_{xy}) \rangle^2}{\langle W({\cal{C}}) \rangle} \ ,
\end{equation}
the FM operator is defined as 
\begin{equation}
\label{FMoperator_defBIS}
\rho \equiv \lim_{R\to\infty} \ H(R,T) \ .
\end{equation} 
Note that this definition differs from the conventional one. In the conventional definition one usually takes $T=R$, so in the limit $R\to\infty$ both temporal and spatial sizes of the Wilson loop and line diverge. 

At finite temperature the size $T$ cannot exceed the temporal lattice extent $N_t$ which is fixed and usually rather small. 
We refer to
such operator as the temporal FM operator. In the present work we compute both
the spatial and the temporal FM operators. The former is defined in a standard
way: one should take the spatial Wilson loop and spatial Wilson line in Eq.~(\ref{FMoperator_def}) and
put $T = R$.

It is not obvious that the temporal FM operator vanishes in the deconfined phase if $T$ is fixed. Below we discuss simple non-rigorous arguments that suggest that this can be the case. 
It is also possible to construct a rigorous proof that any FM operator is vanishing in the deconfined phase of $Z(2)$ theory if $T$ is fixed and the limit is taken as  in~(\ref{FMoperator_defBIS})~\footnote{
  M.P.~Forsstr\" om, private communication.}.

\subsection{FM operator in the region of small Higgs coupling} 

In this subsection we present analytic estimates of the FM operator. Expanding the Higgs part of the action into the character series and keeping only the leading fundamental term at small $\gamma$, the partition function gets the form
\begin{align}
Z &=  \int \prod_l dU(l) \ e^{S_G} \ \int \prod_x dV(x) \ \prod_l \ \left [ 1 + C(\gamma) \ {\rm Tr} \ V^{\dagger}(x) U_n(x) V(x+e_n) \right ]  \nonumber \\
\label{gauge_higgs_pf_ep}
&= Z_G \left [ 1 + \sum_{{\cal{C}}} \ N^{1-P({\cal{C}})} \ 
\left ( C(\gamma) \right )^{P({\cal{C}})} \  \langle W({\cal{C}}) \rangle_G + \cdots  \right ] \ , 
\end{align} 
where $Z_G$ is the partition function of the pure gauge theory and 
the expectation $\langle \cdots \rangle_G$ refers to this pure gauge theory. 
The sum runs over all loops ${\cal C}$ on $\Lambda$ and $P({\cal{C}})$ represents the perimeter of the loop. $C(\gamma)$ is the fundamental coefficient of the $SU(N)$ character expansion. 

The expectation value of the Wilson loop for a certain external loop ${\cal{C}}_{ex}$ 
is
\begin{align}
&\langle W({\cal{C}}_{ex}) \rangle = \langle W({\cal{C}}_{ex}) \rangle_G + 
\sum_{{\cal{C}}} \ N^{1-P({\cal{C}})} \ 
\left ( C(\gamma) \right )^{P({\cal{C}})} \big [ \langle W({\cal{C}}) \  W({\cal{C}}_{ex}) \rangle_G   \nonumber \\
&- \label{wilson_loop_exp_gen} 
\langle W({\cal{C}}) \rangle_G \ \langle W({\cal{C}}_{ex}) \rangle_G \big ] 
+ \cdots  \ = \ \langle W({\cal{C}}_{ex}) \rangle^{(1)}
+ \langle W({\cal{C}}_{ex}) \rangle^{(2)}   \ ,
\end{align} 
where $\langle W({\cal{C}}_{ex}) \rangle^{(1)}$ describes the contribution from the pure gauge theory plus corrections,
\begin{align}
&\langle W({\cal{C}}_{ex}) \rangle^{(1)} \equiv \langle W({\cal{C}}_{ex}) \rangle_G + 
\sum_p \ N^{-3} \ 
\left ( C(\gamma) \right )^4 \big [ \langle {\rm Tr} U_p \  W({\cal{C}}_{ex}) \rangle_G \ \nonumber \\
&- \label{wilson_loop_exp_1} 
\langle {\rm Tr} U_p \rangle_G \ \langle W({\cal{C}}_{ex}) \rangle_G \big ] 
+ \cdots  \ ,
\end{align} 
while $\langle W({\cal{C}}_{ex}) \rangle^{(2)}$ describes the contribution from the loop ${\cal{C}}={\cal{C}}_{ex}$ plus corrections,
\begin{align}
\label{wilson_loop_exp_2}
&\langle W({\cal{C}}_{ex}) \rangle^{(2)} \equiv   N^{1-P({\cal{C}}_{ex})} \ 
\left ( C(\gamma) \right )^{P({\cal{C}}_{ex})}  \big [ \langle  W^2({\cal{C}}_{ex}) \rangle_G - \langle W({\cal{C}}_{ex}) \rangle^2_G \big ] 
+ \cdots  \ .
\end{align} 
Similar expansions for the staple-like Wilson line 
$V({\cal{L}}_{xy})$ read
\begin{align}
\label{wilson_line_exp_gen} 
\langle V({\cal{L}}_{xy}) \rangle  &= \langle V({\cal{L}}_{xy}) \rangle^{(1)} +  \langle V({\cal{L}}_{xy}) \rangle^{(2)} \ ,  \\
\label{wilson_line_exp_1} 
\langle V({\cal{L}}_{xy}) \rangle^{(1)} &= N \left [ \frac{C(\gamma)}{N} \right ]^{R+T} \  \biggl [ 1 + \left [ \frac{C(\gamma)}{N} \right ]^2 \ 
\sum_{p\in {\cal{L}}_{xy}} \ \langle {\rm Tr} U_p \rangle_G + \cdots \biggr ] 
\ ,   \\
\label{wilson_line_exp_2} 
\langle V({\cal{L}}_{xy}) \rangle^{(2)} &= N \sum_{{\cal{M}}_{xy}} \ 
\left [ \frac{C(\gamma)}{N} \right ]^{| {\cal{M}}_{xy} |} \ 
\langle W({\cal{L}}_{xy} \cup {\cal{M}}_{xy}) \rangle_G  \ .
\end{align}
The first term in $\langle V({\cal{L}}_{xy}) \rangle^{(1)}$ arises when all links from the Wilson line are covered by links
taken from the gauge-Higgs action. The second term describes the first correction, in which the sum runs over all plaquettes that have a link in common with the Wilson line. The sum in $\langle V({\cal{L}}_{xy}) \rangle^{(2)}$ runs over all paths connecting points $x$ and $y$ and the Wilson loop $W({\cal{L}}_{xy} \cup {\cal{M}}_{xy})$ is constructed from the Wilson line ${\cal{L}}_{xy}$ whose endpoints are connected by the path ${\cal{M}}_{xy}$.
These expressions are valid for $Z(N)$, $U(1)$ and $SU(N)$ models under the assumption that all series converge if $\gamma$ is small. 
For $SU(2)$ one has $C(\gamma)/N=\frac{1}{2} \frac{I_2(\gamma)}{I_1(\gamma)}$, where $I_1,I_2$ are modified Bessel functions of the first kind.

{\it Confinement phase, small $\beta$}. 
In this region the Wilson loop in the pure gauge theory decays with the area law 
$\langle W({\cal{C}}_{ex}) \rangle_G\approx (C(\beta))^{RT}$. 
Combining (\ref{wilson_loop_exp_1}) and (\ref{wilson_loop_exp_2}) yields
\begin{eqnarray}
\label{wilson loop_z2_conf}
\langle W({\cal{C}}_{ex}) \rangle = (C(\gamma))^{2R + 2T} 
\left [ 1 - (C(\beta))^{2RT}  \right ] + (C(\beta))^{RT} + \cdots \ . 
\end{eqnarray}
Combining (\ref{wilson_line_exp_1}) and (\ref{wilson_line_exp_2}) produces
\begin{eqnarray}
\label{wilson line_z2_conf}
\langle V({\cal{L}}_{xy}) \rangle  &=& (C(\gamma))^{R + T} +  
\sum_{{\cal{M}}_{xy}} \ 
( C(\gamma) )^{| {\cal{M}}_{xy} |} \ 
\langle W({\cal{L}}_{xy} \cup {\cal{M}}_{xy}) \rangle_G  \\
&\approx& (C(\gamma))^{R + T} +  (C(\gamma))^{R} \ (C(\beta))^{\frac{RT}{2}}  + \cdots   \ .  \nonumber 
\end{eqnarray}
For any FM operator, spatial or temporal, it gives 
\begin{equation}
\label{FM_z2_conf} 
H(R,T) = \frac{\left ( (C(\gamma))^{R + T} +  (C(\gamma))^{R} \ (C(\beta))^{\frac{RT}{2}}  \right )^2}{(C(\gamma))^{2R + 2T} 
 + (C(\beta))^{RT}}  \ . 
\end{equation}
The ratio goes to a constant in the limit $R\to\infty$, if $T>2$ is fixed.

{\it Deconfinement phase, large $\beta$: temporal FM operator}. 
In this region, the temporal Wilson loop in  the pure gauge theory decays with a perimeter law 
$\langle W({\cal{C}}) \rangle_G\approx \exp\left ( -m(\beta) P({\cal{C}}) \right )$~\cite{Borgs_Seiler},
\begin{eqnarray}
\label{wilson loop_z2_deconf}
\langle W({\cal{C}}_{ex}) \rangle = \exp\left ( -m(\beta) (2R + 2T) \right ) 
+ (C(\gamma))^{2R + 2T} \ .
\end{eqnarray}
By combining (\ref{wilson_line_exp_1}) and (\ref{wilson_line_exp_2}) we have 
\begin{eqnarray}
\label{wilson line_z2_deconf}
\langle V({\cal{L}}_{xy}) \rangle  = (C(\gamma))^{R + T} + 
\sum_{{\cal{M}}_{xy}} \ 
\left ( C(\gamma) \right )^{| {\cal{M}}_{xy} |} \ 
\exp\left ( -m(\beta) (R + T + | {\cal{M}}_{xy} | ) \right ) \ ,
\end{eqnarray}
where $|{\cal M}_{xy}|$ is the length of the path ${\cal M}_{xy}$.

For the temporal FM operator it gives 
\begin{equation}
\label{FM_z2_deconf} 
H(R,T) = \frac{\left ( (C(\gamma))^{R + T} +  e^{-m(\beta) (R + T)} \ 
\sum_{{\cal{M}}_{xy}} \ 
\left ( C(\gamma) \ e^{-m(\beta)} \right )^{| {\cal{M}}_{xy} |}  \right )^2}{(C(\gamma))^{2R + 2T} + e^{-m(\beta) (2R + 2T)}} \ . 
\end{equation}
The leading contribution comes from the term $| {\cal{M}}_{xy} |=R$,
\begin{equation}
\label{FM_z2_deconf_1} 
H(R,T) = (C(\gamma))^{2R} \ \frac{\left ( (C(\gamma))^T +  e^{-m(\beta) (2R + T)}
  \right)^2}{(C(\gamma))^{2R + 2T} + e^{-m(\beta) (2R + 2T)}} \ .
\end{equation}
There is a competition between two terms in the denominator. The first term arises due to the screening from the matter fields, while the second one arises from gluonic screening. Although we do not know how to reliably calculate the coefficient $m(\beta)$ in the perimeter law, at least it is expected that $m(\beta)\sim {\cal{O}}(1/\beta)$ as this is so in perturbation theory. 
Therefore, if $\beta$ is sufficiently large the gluonic screening dominates
and we obtain  
\begin{equation}
\label{FM_z2_deconf_final} 
H(R,T) \approx  \left ( C(\gamma) \ e^{m(\beta)} \right )^{2R + 2T} \ .
\end{equation}
Even if $T$ is fixed, $H(R,T)\to 0$ in the limit $R\to\infty$.

{\it Deconfinement phase, large $\beta$: spatial FM operator}. 
In this region the spatial Wilson loop in the pure gauge theory decays with an area law. Taking $T=R$ leads to the conclusion that the spatial FM operator tends to a constant, pretty much like the behavior that appears in the confinement phase.

\section{Numerical results}
\label{num_results}

In this section we present the results obtained with Monte Carlo simulations for the various types of overlap and FM operators. Gauge and Higgs fields have been updated by use of the classical Heat-Bath algorithm \cite{HB}, improved with overrelax \cite{overrelax}. 

%On the other hand, since the Higgs phase %behaves in a sense like a spin-glass %phase, also typical methods to speed up %the Monte Carlo process for spin-glasses %have been included in our computer codes. 
Initially, we have used the tempered Monte Carlo \cite{tempered} and averaging over replicas \cite{edwards_75}, which were helpful in the $Z(2)$ model studied in~\cite{z2_gauge_Higgs}. However, it turned out that these techniques are not so relevant for the study of the $SU(2)$ gauge-Higgs theory, at least for the volumes considered here. This is because (i) the 
Heat-Bath algorithm improved with overrelax hits was enough to obtain a satisfactory decorrelation rate, and (ii) the variability entailed by replicas stayed well within the error bars or stated diversely, that variability was insignificant. Therefore, we decided to abandon them. 

\subsection{The overlap operator}
\label{overlap_num_res}

\begin{figure}[H]
\centering
\includegraphics[width=0.45\linewidth,clip]
{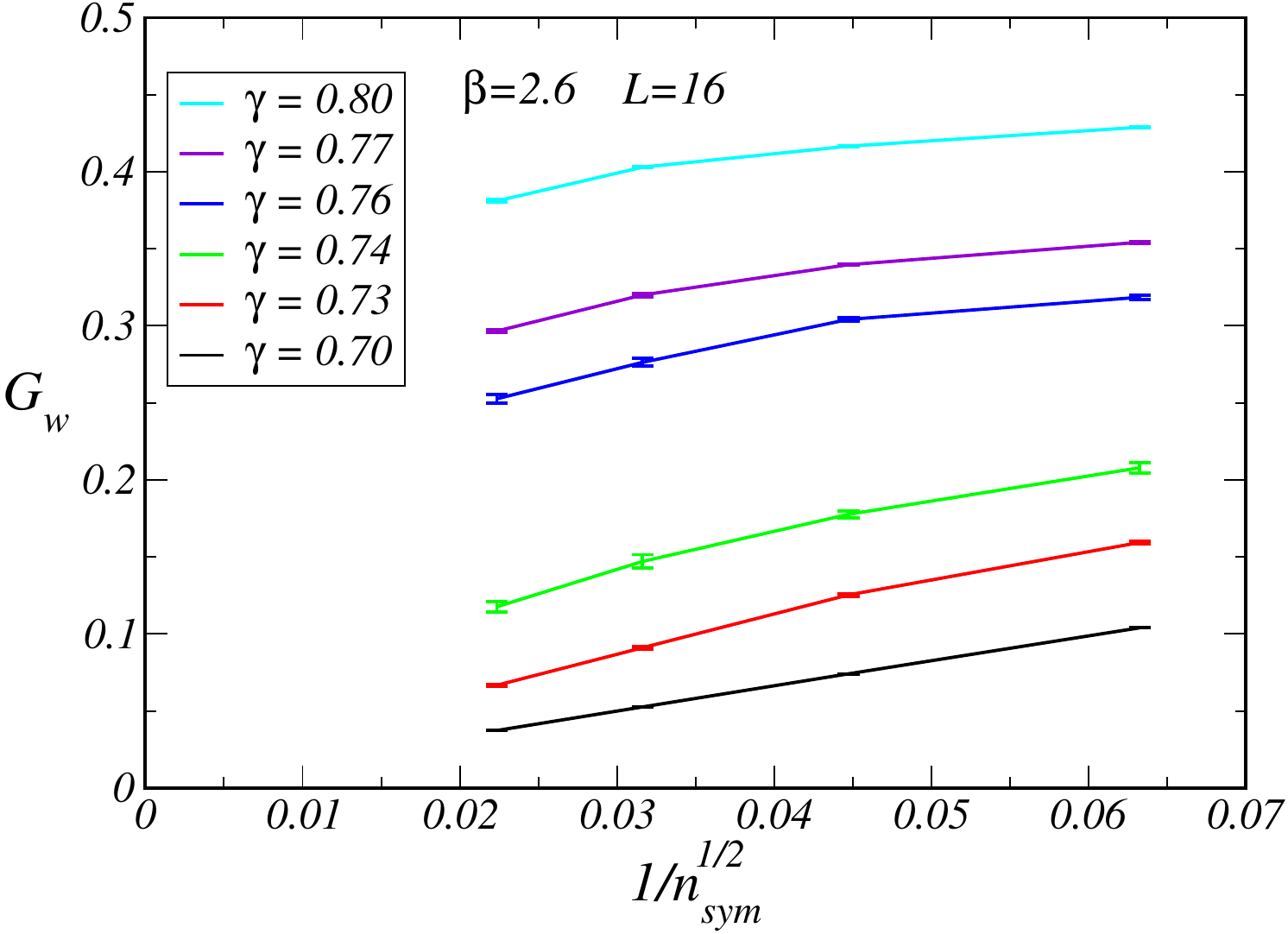}
\includegraphics[width=0.45\linewidth,clip]
{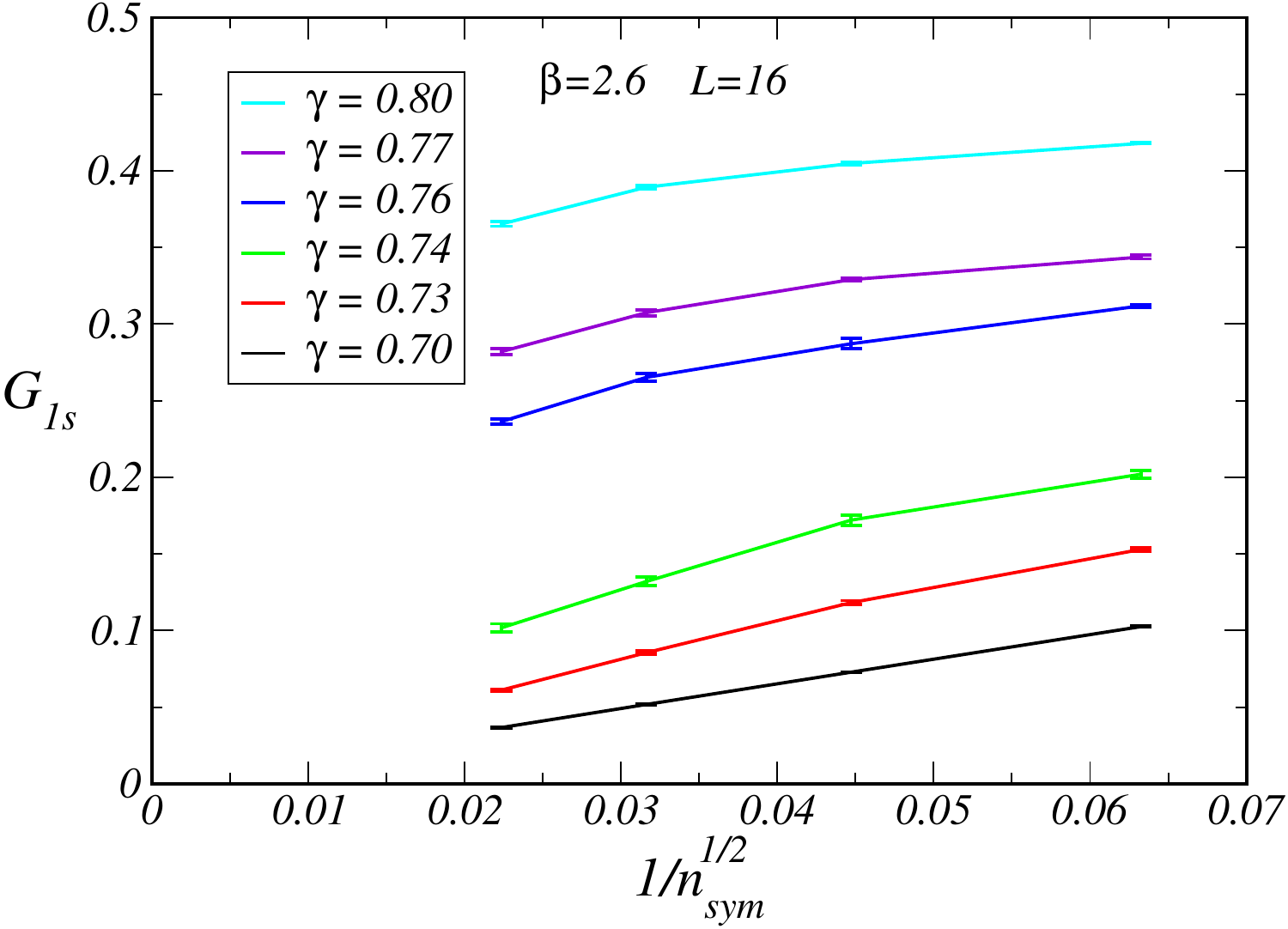}
\caption{Overlap operators at $\beta=2.6$, $N_t=4$, $L=16$.}
\label{fig:overlap.b2.6.L16}
\end{figure}

\begin{figure}[H]
\centering
\includegraphics[width=0.45\linewidth,clip]
{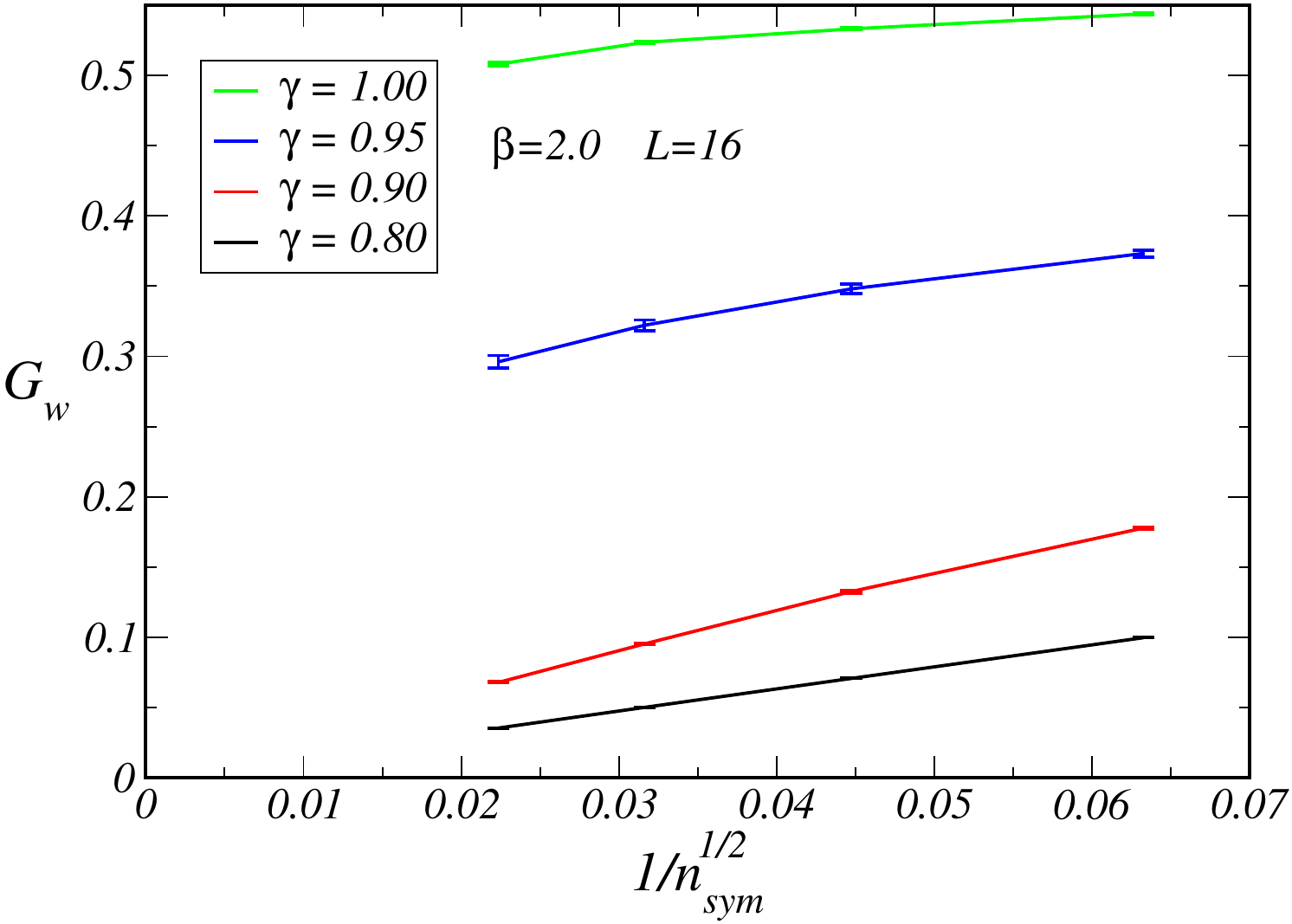}
\includegraphics[width=0.45\linewidth,clip]
{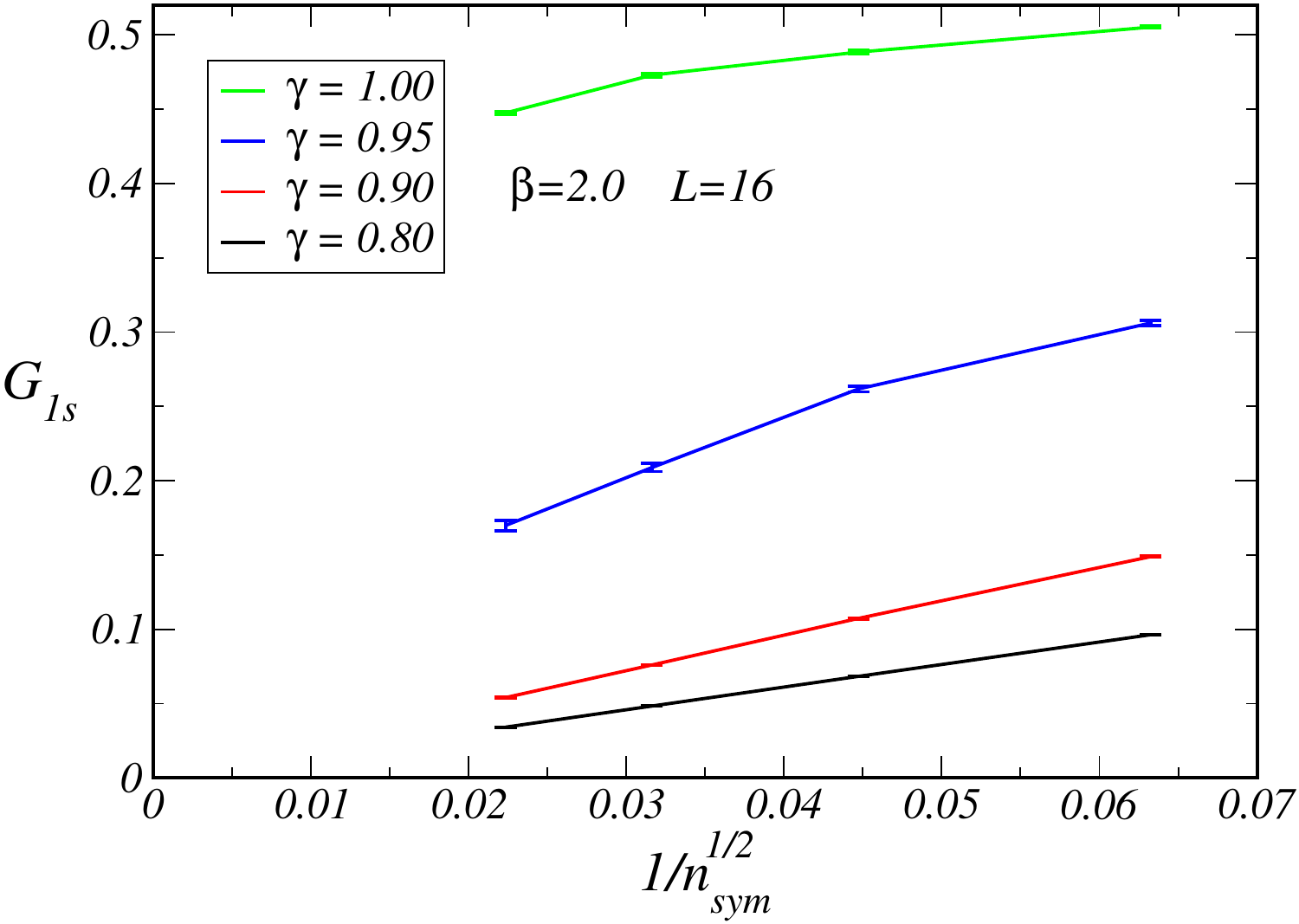}
\caption{Overlap operators at $\beta=2.0$, $N_t=4$, $L=16$.}
\label{fig:overlap.b2.0.L16}
\end{figure}

\begin{figure}[H]
\centering
\includegraphics[width=0.45\linewidth,clip]
{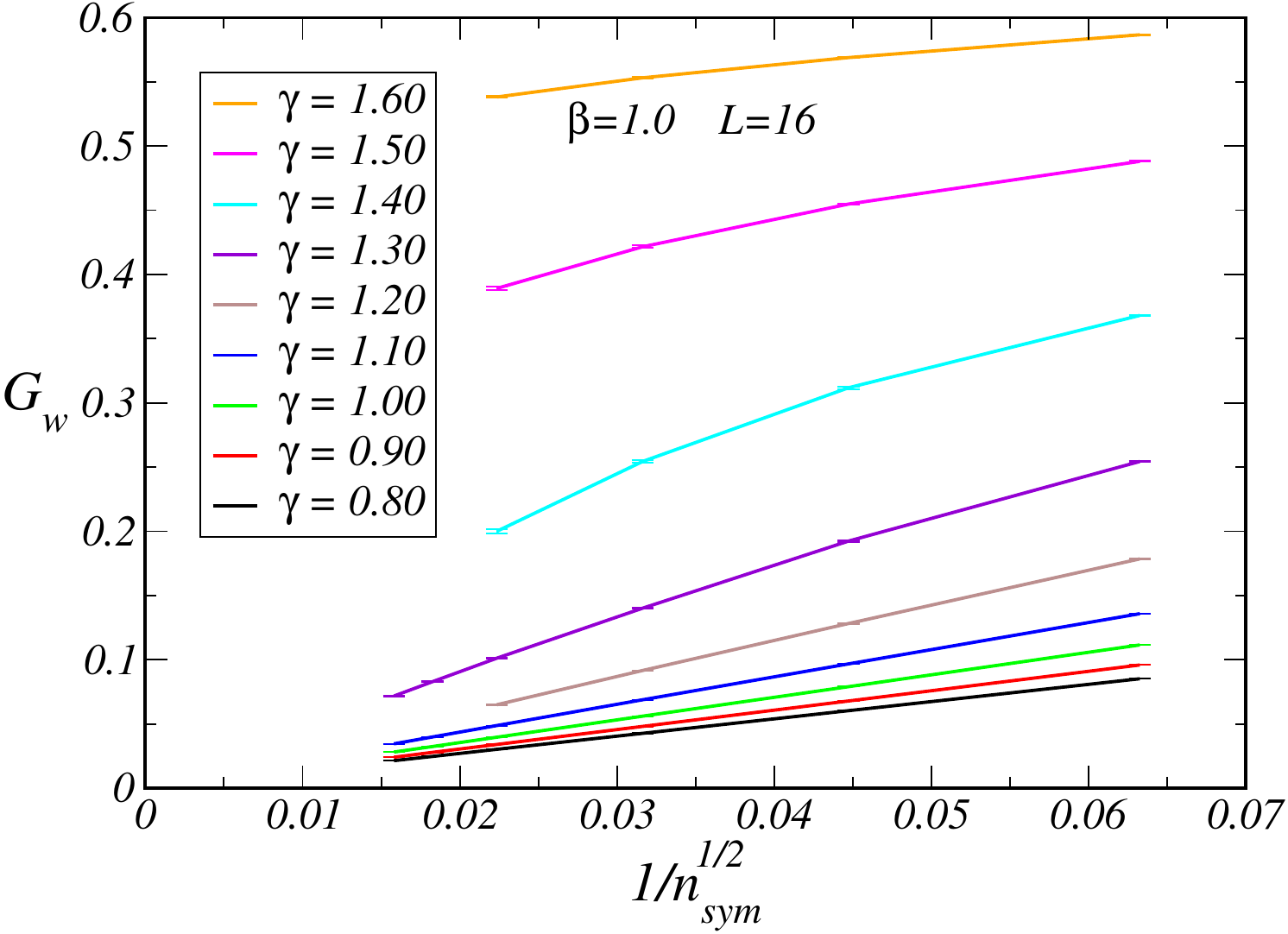}
\includegraphics[width=0.45\linewidth,clip]
{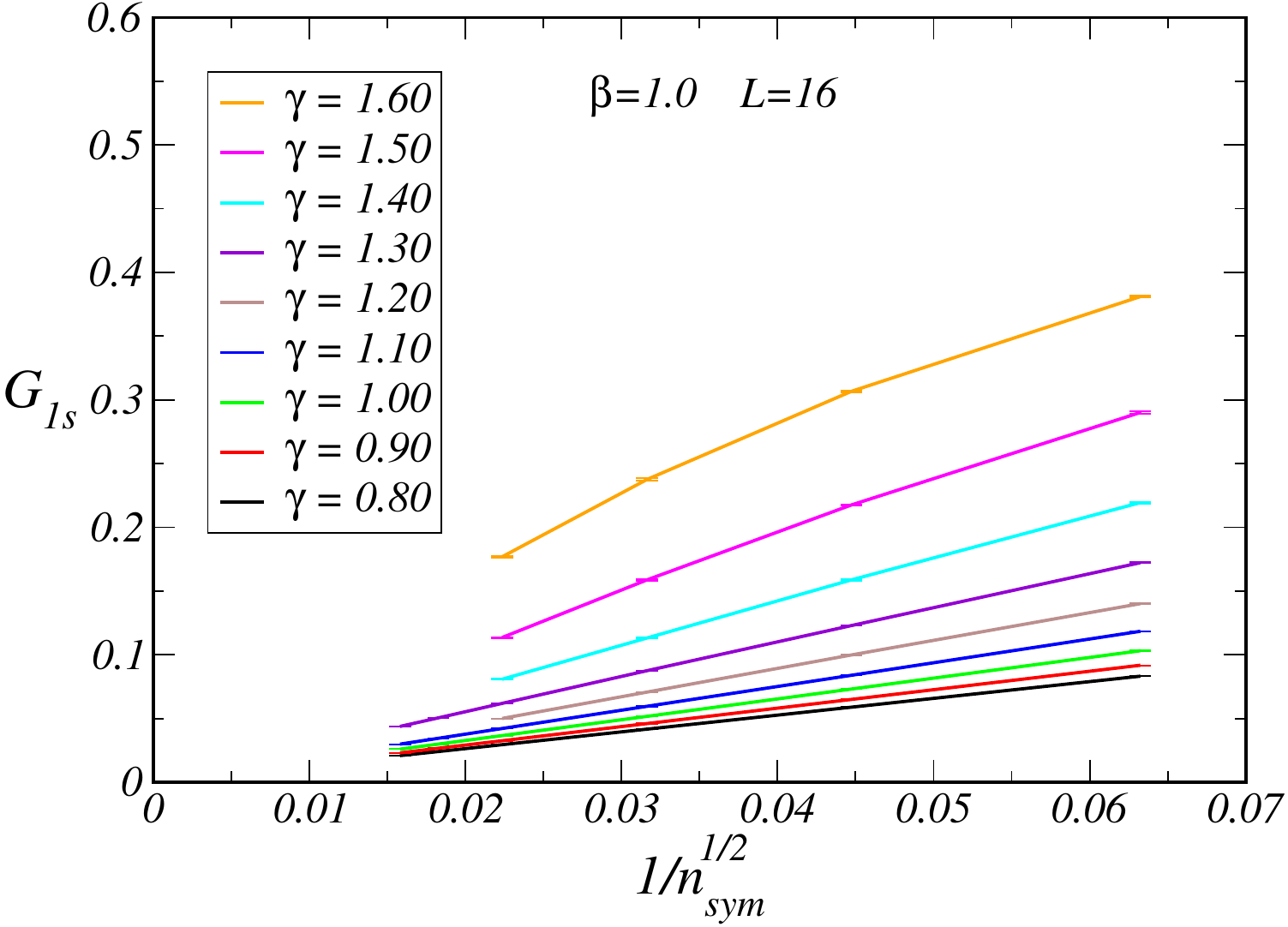}
\caption{Overlap operators at $\beta=1.0$, $N_t=4$, $L=16$.}
\label{fig:overlap.b1.0.L16}
\end{figure}

The overlap operator can distinguish between confinement and Higgs phases 
as well as between deconfined and Higgs phases. It is expected that this operator decays to zero in confinement and deconfined phases, while it tends to a constant in the Higgs phase, where the global $SU_l(2)$ symmetry acting on the Higgs fields is spontaneously broken. According to this scenario, the best strategy is to fix the gauge coupling $\beta$ and perform simulations for several Higgs couplings $\gamma$ in the vicinity of the critical line. 
We performed such simulations for many $\beta$ values in the range 
$\beta\in [0.5 - 2.6]$ on the lattice with temporal extent $N_t=4$ 
and spatial size $L=16$. The maximum value of $n_{\rm sym}$ was $n_{\rm sym}=2000$ in all cases. Both the $G_{w}$ and $G_{1s}$ operators were computed. The typical behavior of the results in all cases is well represented by Fig.\ref{fig:overlap.b2.6.L16} - Fig.\ref{fig:overlap.b1.0.L16} as a function of $1/\sqrt{n_{\rm sym}}$. Data were taken by independent simulations for each of the four or five values of $n_{\rm sym}$ shown in the figures. Lines joining the resulting points are drawn to guide the eye.

Fig.\ref{fig:overlap.b2.6.L16} shows the overlap operators in the deconfinement/Higgs region. Fig.\ref{fig:overlap.b2.0.L16} presents the overlap  in the confinement/Higgs region for $\beta=2.0$, where, presumably, the thermodynamic transition takes place. Finally, Fig.\ref{fig:overlap.b1.0.L16} shows the overlap in the confinement/Higgs region for $\beta=1.0$ corresponding to a crossover transition between the two phases. 

\begin{figure}[H]
\centering
\includegraphics[width=0.45\linewidth,clip]
{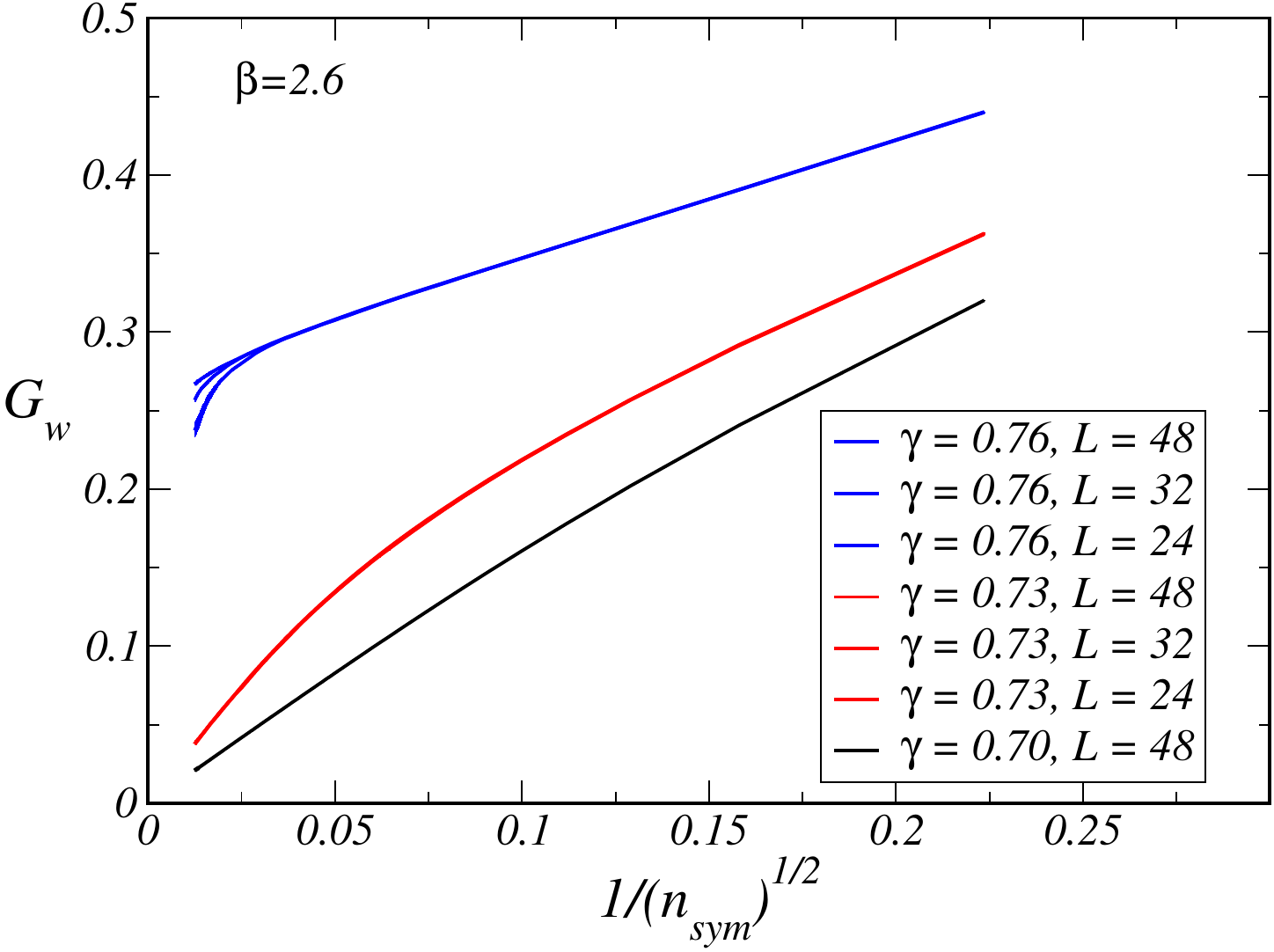}
\includegraphics[width=0.45\linewidth,clip]
{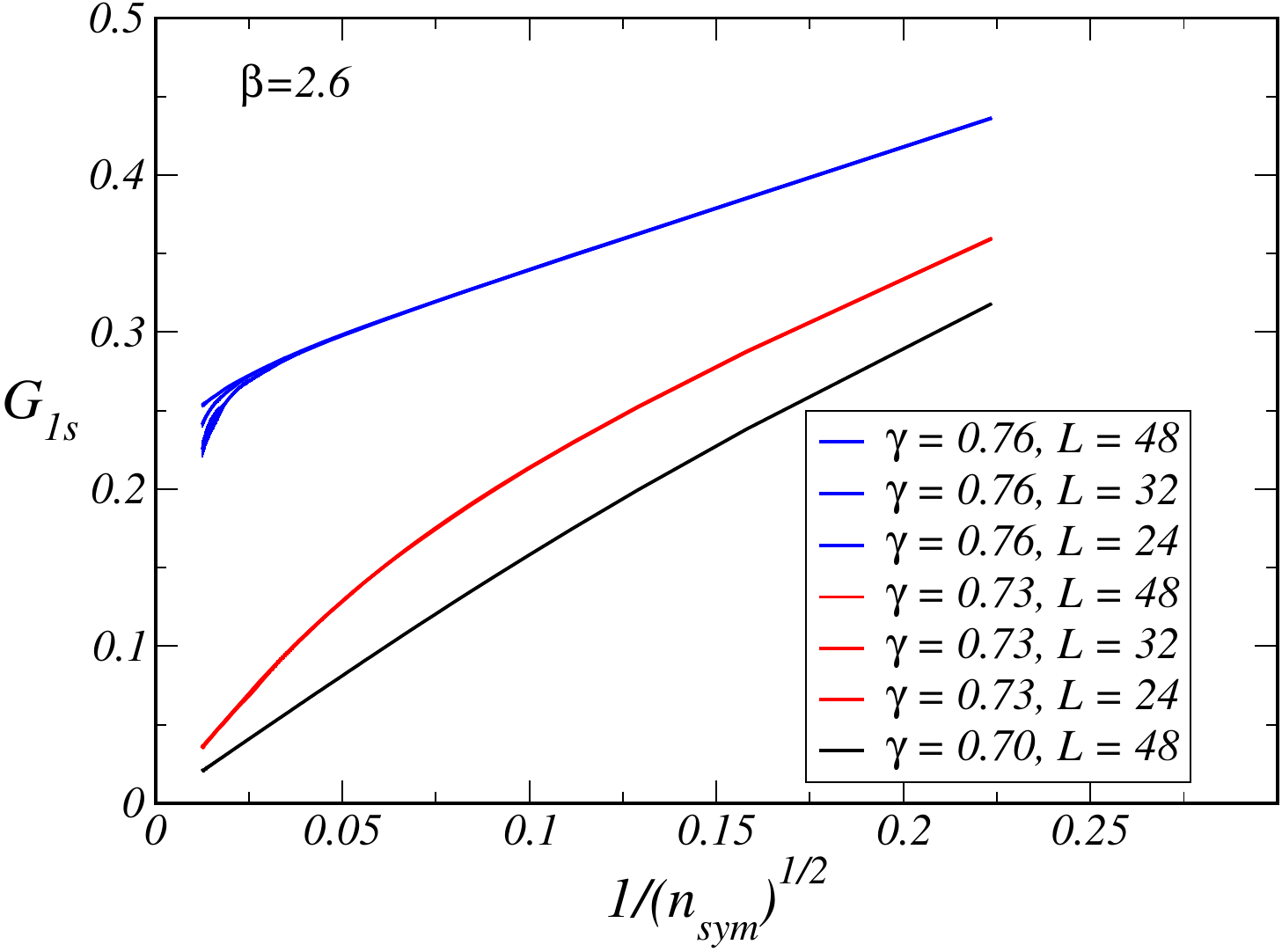}
\caption{Overlap operators at $\beta=2.6$, $N_t=4$.}
\label{fig:overlap.b2.6.TL}
\end{figure}

\begin{figure}[H]
\centering
\includegraphics[width=0.45\linewidth,clip]
{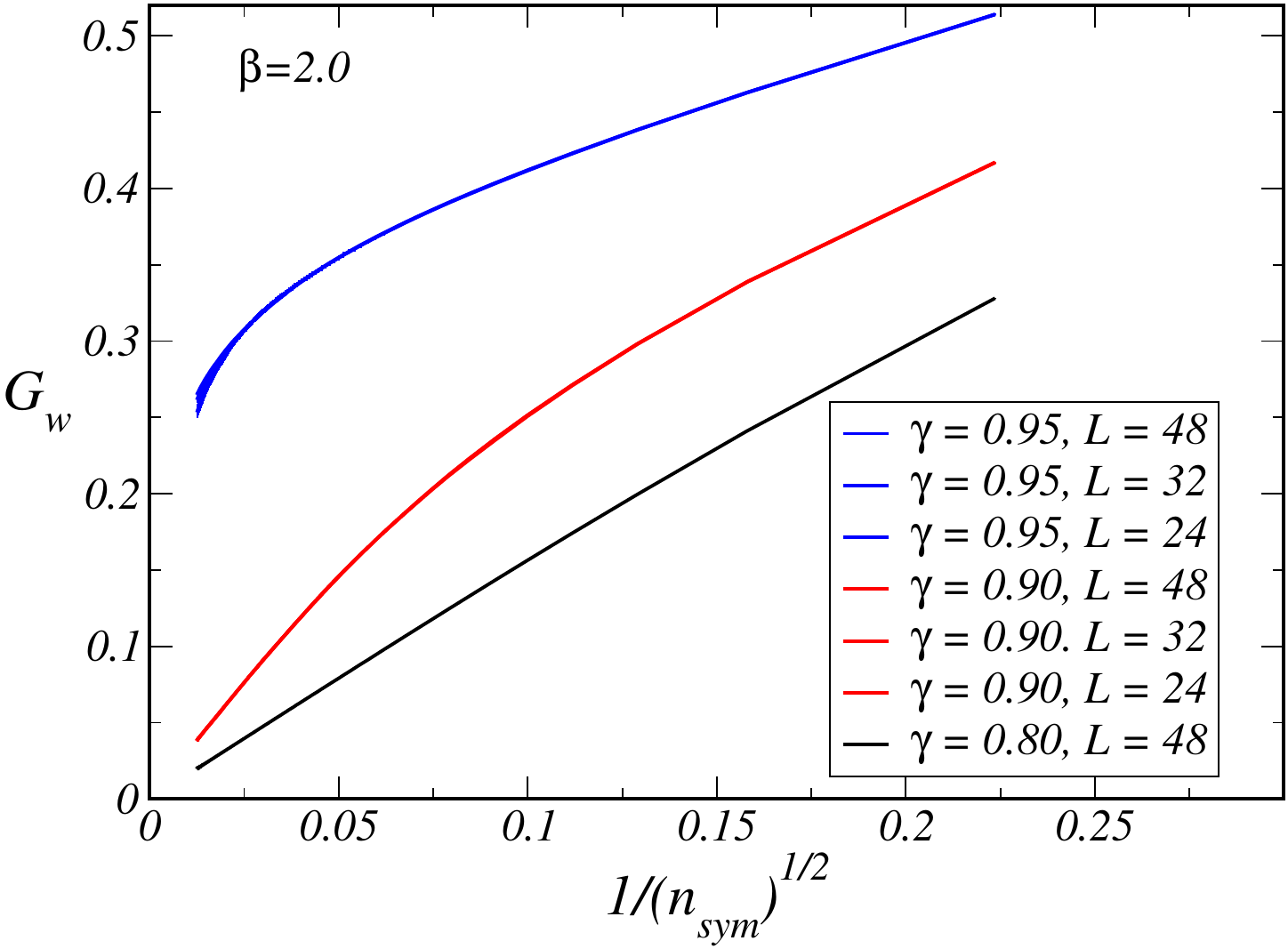}
\includegraphics[width=0.45\linewidth,clip]
{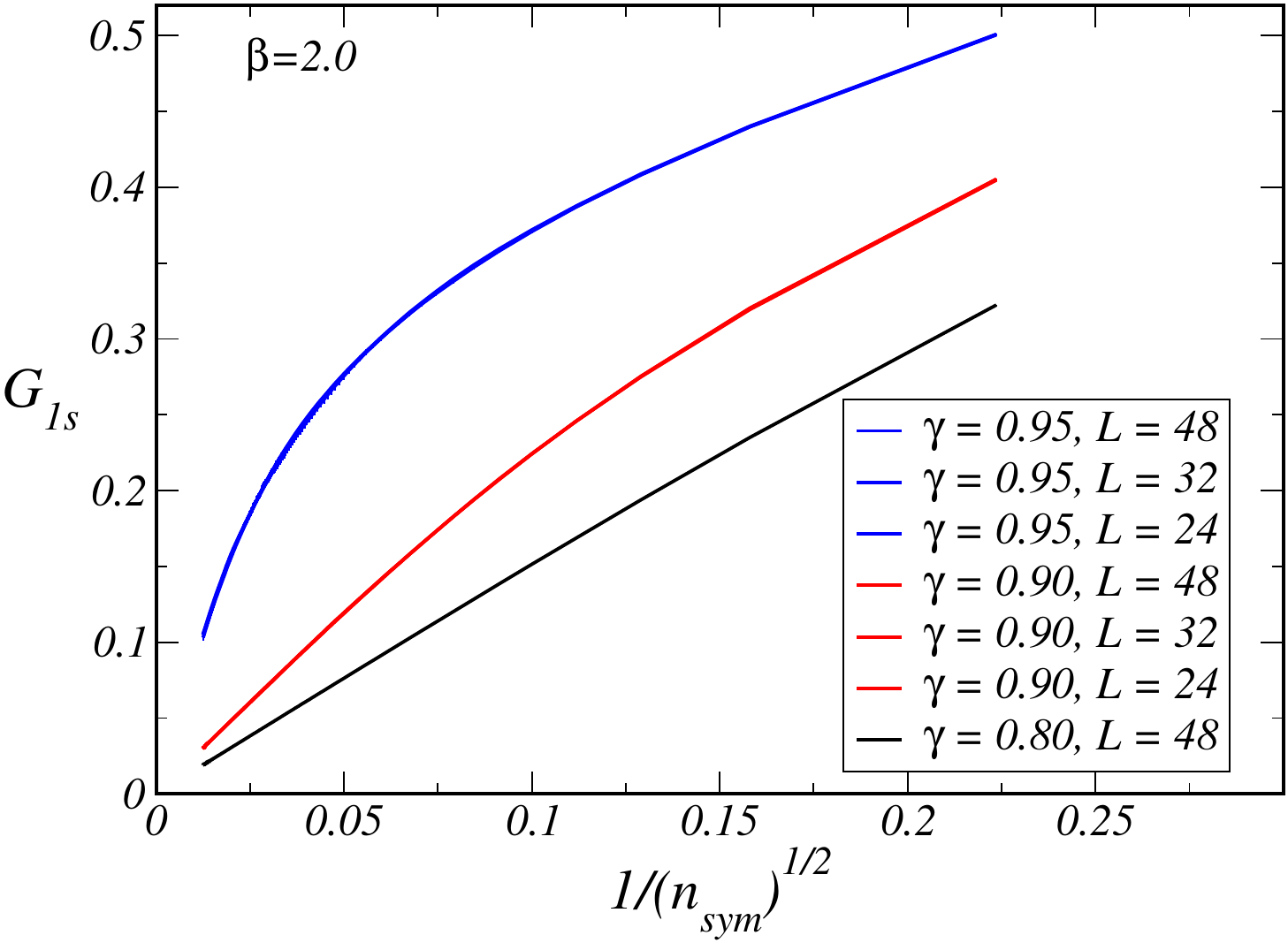}
\caption{Overlap operators at $\beta=2.0$, $N_t=4$.}
\label{fig:overlap.b2.0.TL}
\end{figure}

The correct way to evaluate the overlap operator is to introduce first a small external gauge-invariant perturbation as described in~\cite{greensite_22_overlap}. Then, one has to take the thermodynamic limit followed by the limit of vanishing perturbation. Such a strategy, however, is hardly feasible in realistic simulations. 
Therefore, to check the consistency of the results, we repeated the above simulations for $\beta=2.6$ and $\beta=2.0$ by using several increasingly large lattice volumes $L=24,32,48$, and by single runs where $n_{\rm sym}$ varies from~1 to~6400. The corresponding results are presented on Fig.\ref{fig:overlap.b2.6.TL} and Fig.\ref{fig:overlap.b2.0.TL}, as usual as a function of $1/\sqrt{n_{\rm sym}}$. Error bars in these plots were removed because they were as small as the very thickness of the lines.

If for a certain pair of couplings $(\beta,\gamma)$ the results of the overlap point to zero at large $n_{\rm sym}$ for all $L$, then it is
deemed that the overlap is zero for those values of the couplings. Instead, if by increasing $L$ the results tend to bend upwards, thus deviating from zero, as it occurs for $\gamma=0.76$ in Fig.\ref{fig:overlap.b2.6.TL} and a bit less for $\gamma=0.95$ in Fig.\ref{fig:overlap.b2.0.TL}, then we consider that the overlap is non-vanishing at those values of $\beta,\gamma$.

%The inspection of Figs.\ref{fig:overlap.b2.6.L16}-\ref{fig:overlap.b1.0.L16}, as well as Figs.\ref{fig:overlap.b2.6.TL}, \ref{fig:overlap.b2.0.TL},
%  shows that, for each $\beta$, there is a region of small
%  values of $\gamma$ where the overlap operators behave linearly with
%  $1/\sqrt{n_{\rm sym}}$ and extrapolate to zero in the limit of infinite
%$n_{\rm sym}$. For larger values of $\gamma$ the linear behavior is lost and
%the overlap operators extrapolate to nonvanishing values. 
Thus, either the overlap tends to zero or not. The value 
of $\gamma$ separating these two different regimes is, by definition, the
critical one, $\gamma_{\rm c}$. At each given $\beta$, we tried to determine
$\gamma_{\rm c}$ by fitting each $\gamma$ data set and looking for the
  $n_{\rm sym} \to \infty$ extrapolation. However, the systematics of the
fitting procedure, namely the extension of the $n_{\rm sym}$ region to
be considered and the 
choice of the fitting function, prevented us from reaching precise
determinations of $\gamma_{\rm c}$.  We therefore limit ourselves to a
(rather conservative) determination of a 1-$\sigma$ error band of $\beta$-dependent values
of $\gamma_{\rm c}$,  as extracted from data for both $G_w$ and $G_{1s}$, which
  are displayed in Fig.\ref{fig:phase diagram_overlap}. The upper and lower limits of the bands were depicted by extrapolation of data. The two plots seem to be in fair agreement.

\subsection{The FM operator}
\label{FM_num_res}

The FM operator was computed on the lattice with temporal extent $N_t=8$ 
and spatial extent $L=32$. Three values of $\beta$ were used in the simulations: 
$\beta=2.0$ for the confinement/Higgs phase and $\beta=2.9$, $\beta=2.6$ for the deconfined/Higgs phase. The typical number of measurements was $200$ K. 
In the confinement/Higgs phases both temporal and spatial operators tend to 
a constant that decreases with $\gamma$. In fact, in the confinement region 
$\gamma<0.95$ this constant becomes very small. Despite the large statistics, error bars are rather big and this fact prevented us from obtaining a reliable determination of the constant in this region.

\begin{figure}[H]
\centering
\includegraphics[width=0.47\linewidth,clip]
{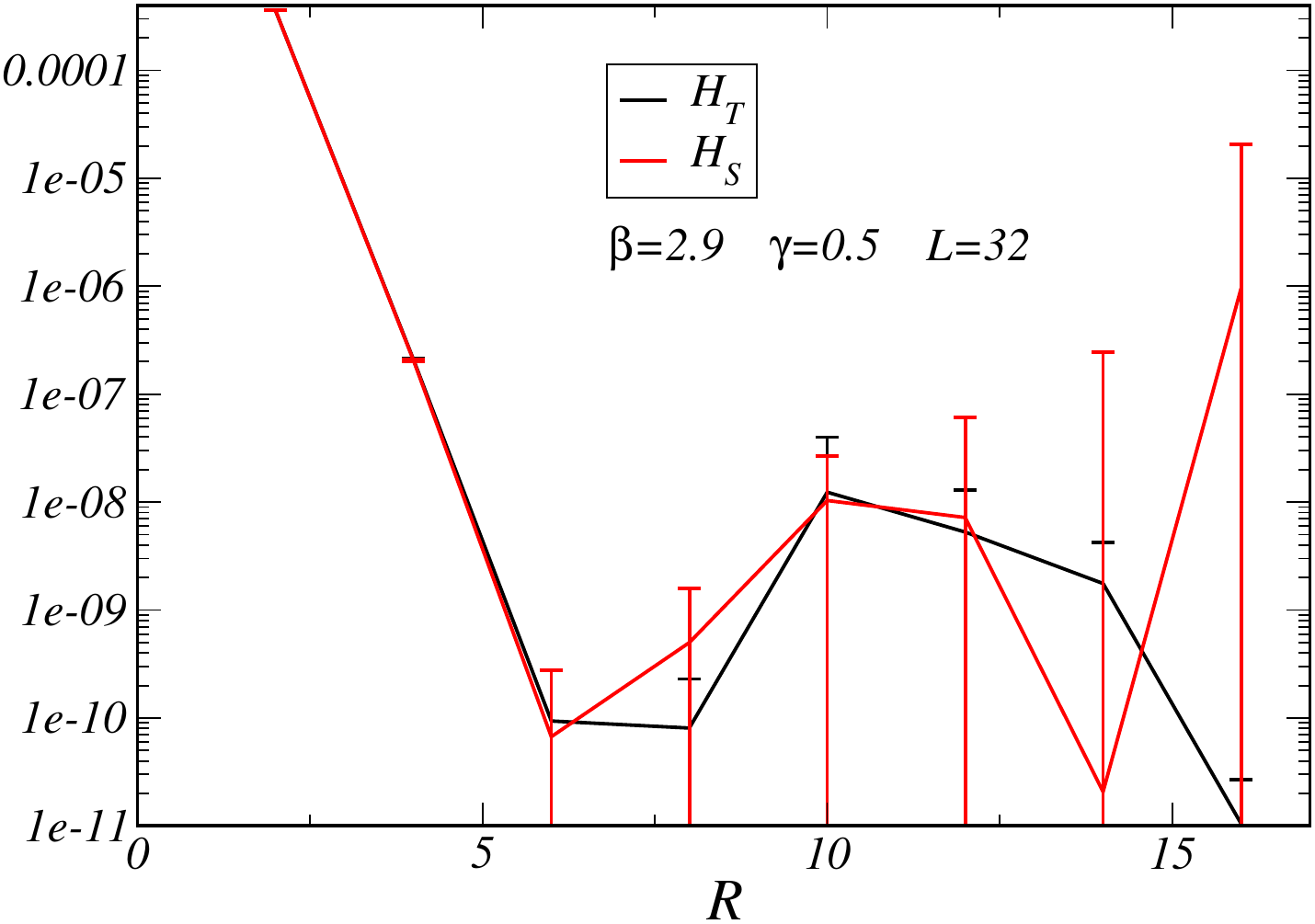}
\includegraphics[width=0.445\linewidth,clip]
{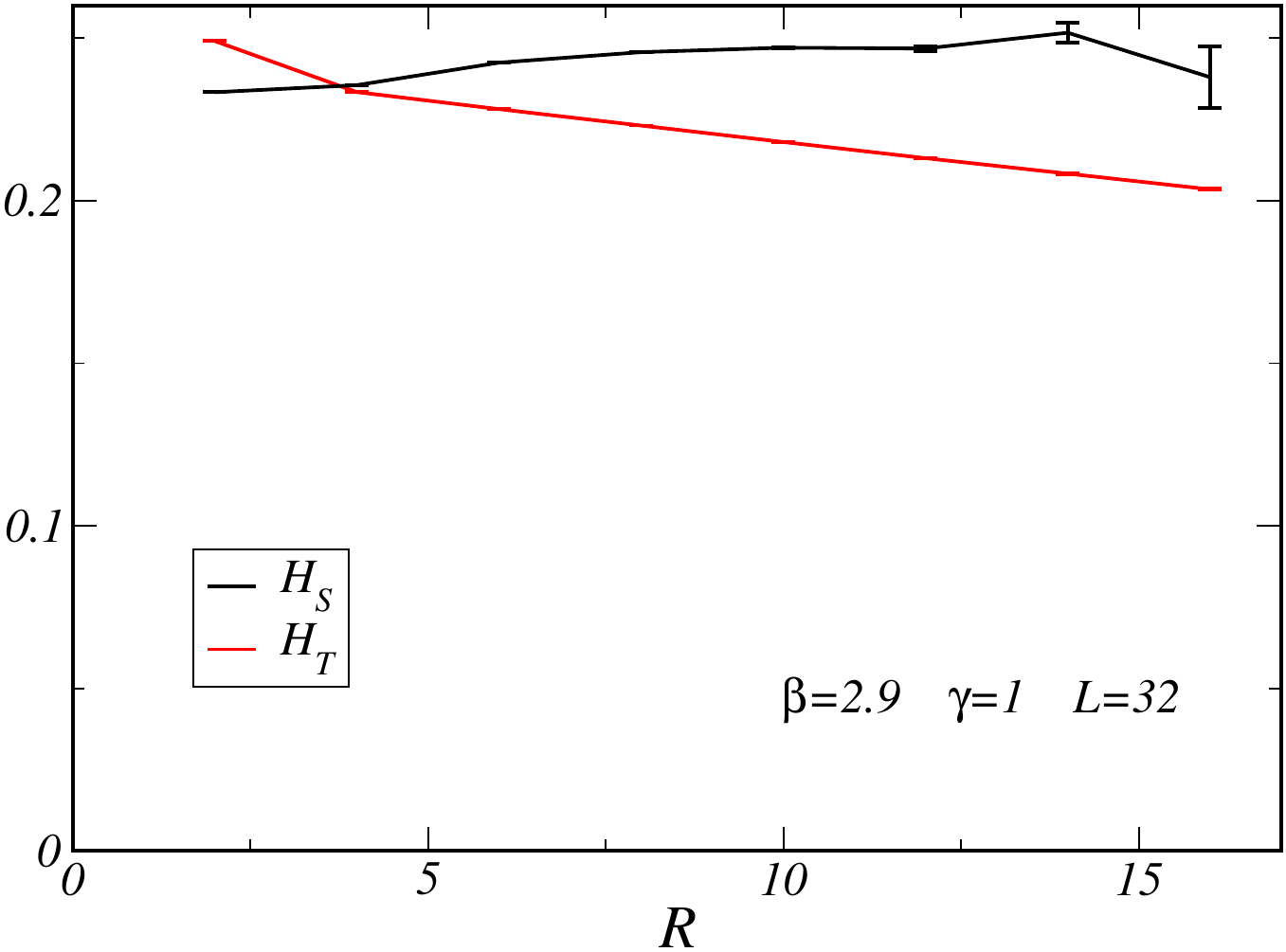}
\caption{The FM operator in the deconfinement-Higgs region, $\beta=2.9$.}
\label{fig:fm_b2.9}
\end{figure}

\begin{figure}[H]
\centering
\includegraphics[width=0.30\linewidth,clip]
{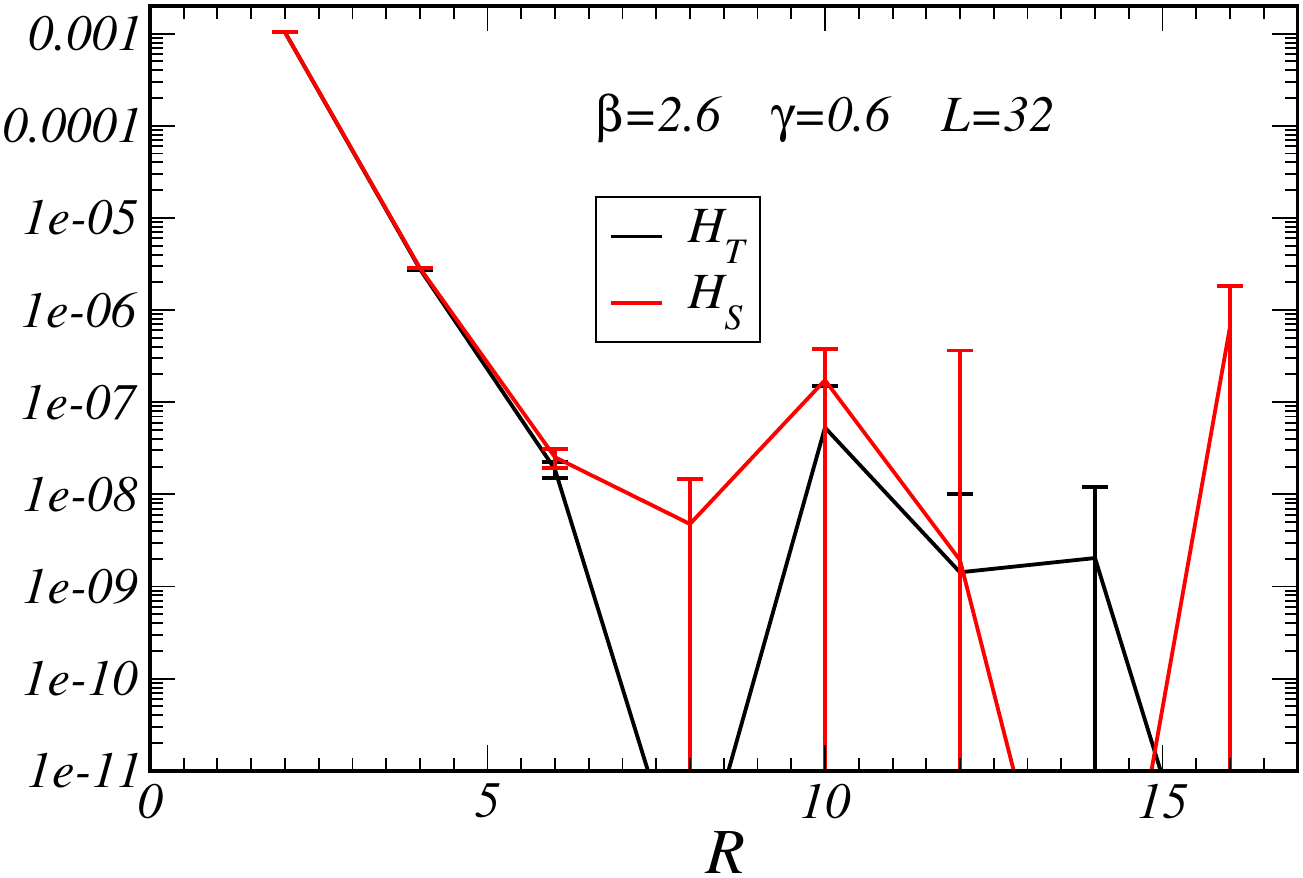}
\includegraphics[width=0.30\linewidth,clip]
{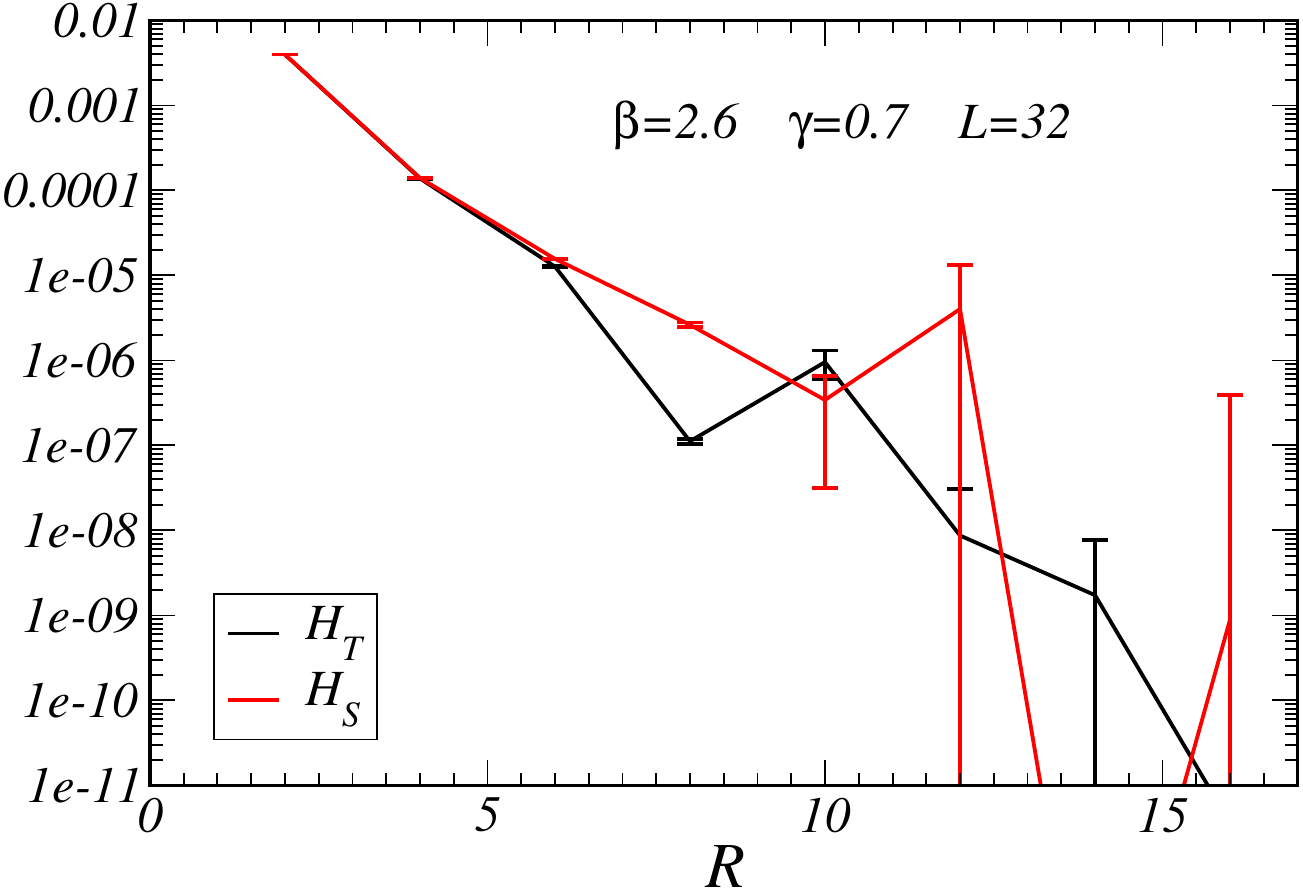}
\includegraphics[width=0.30\linewidth,clip]
{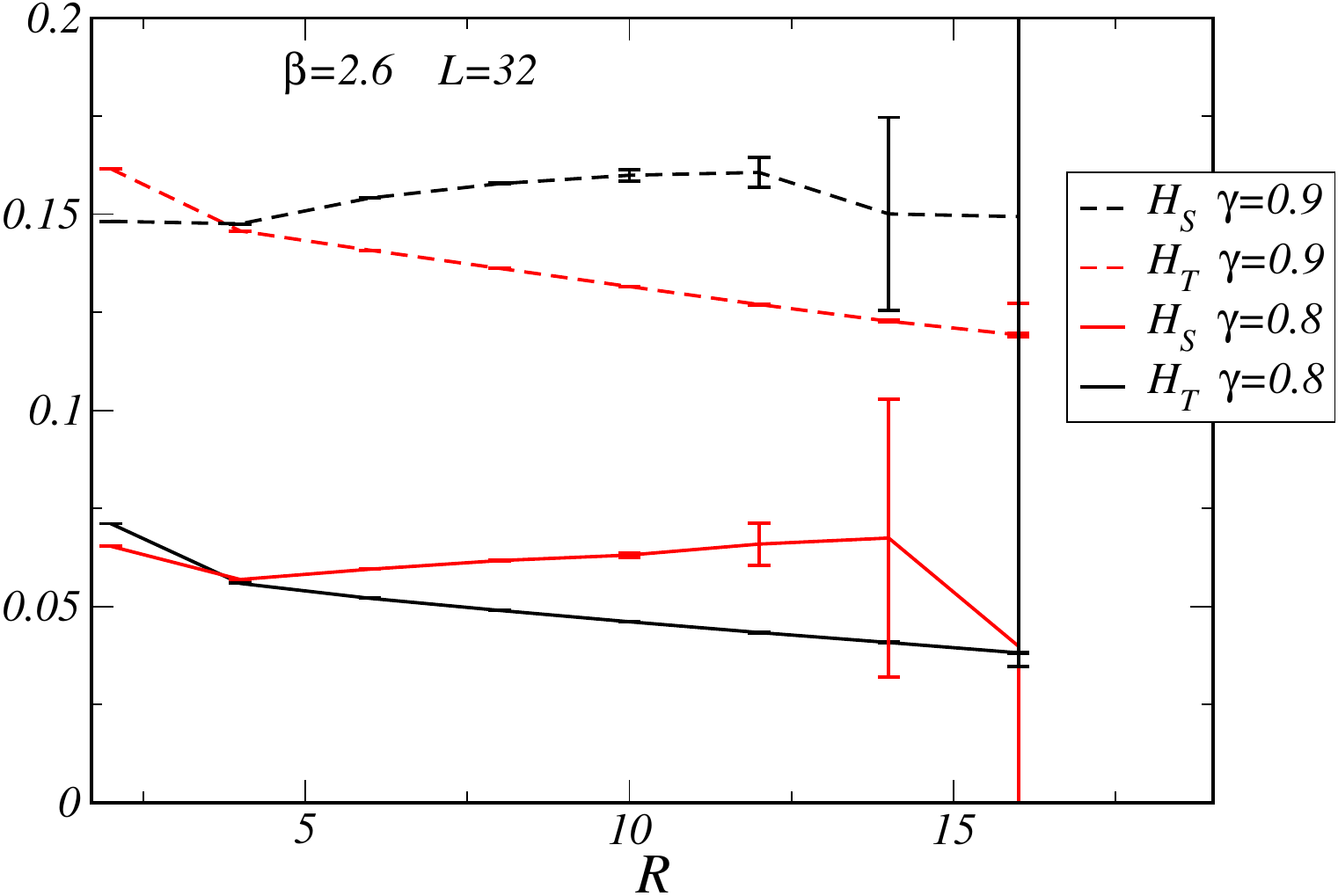}
\caption{The FM operators in the deconfinement-Higgs region, $\beta=2.6$.}
\label{fig:fm_b2.6}
\end{figure}

The most interesting region is, of course, the deconfinement region. 
Results in this region for $\beta=2.9$, $\beta=2.6$ are shown on Fig.\ref{fig:fm_b2.9} and Fig.\ref{fig:fm_b2.6}. We denote by $H_T$ ($H_S$) the temporal (spatial) operator. For the temporal operator, we use $T=N_t/2$.
In the Higgs phase, corresponding to $(\beta=2.9, \gamma=1)$ and $(\beta=2.6, \gamma=0.8, 0.9)$, both FM operators converge to a constant (right plots 
on Figs.\ref{fig:fm_b2.9} - \ref{fig:fm_b2.6}). Instead, in the deconfined phase both operators seem to vanish with distance increasing (left plot on Figs.\ref{fig:fm_b2.9} and left and middle plots on Figs.\ref{fig:fm_b2.6}). Notice the use of the logarithmic scale for the vertical axis in these latter plots.
Large error bars at large distances prevent us from making a solid conclusion 
regarding the vanishing of the two FM operators.  

\section{Summary} 
\label{summary}

In this paper we studied two types of operators in order to understand 
if they can provide order parameters able to reveal the presence of phase transitions in the finite-temperature phase diagram of gauge theories with dynamical matter fields. Concretely, we chose the $SU(2)$ gauge-Higgs LGT in $(3+1)$ dimensions due to its non-trivial phase structure. In effect, it exhibits three different phases ---confined, Higgs and deconfined--- depending on the values of the gauge and Higgs couplings. 
One of these operators is the overlap proposed by Greensite and Matsuyama to distinguish confinement and Higgs phases. Its construction is based on the analogy between the Higgs part of the gauge-Higgs action and spin-glass Hamiltonians. The second operator is the Fredenhagen-Marcu operator which can distinguish the deconfined from the Higgs phases of the theory. Both operators have been tested in zero-temperature gauge-Higgs systems with $Z(2)$ and $SU(2)$ gauge groups. It was demonstrated that they can indeed serve as order parameters of the corresponding phase transitions. 
Our main conclusions regarding the behavior of these operators at finite temperature can be summarized as follows.  

\begin{figure}[H]
\centering
\includegraphics[width=0.47\linewidth,clip]
{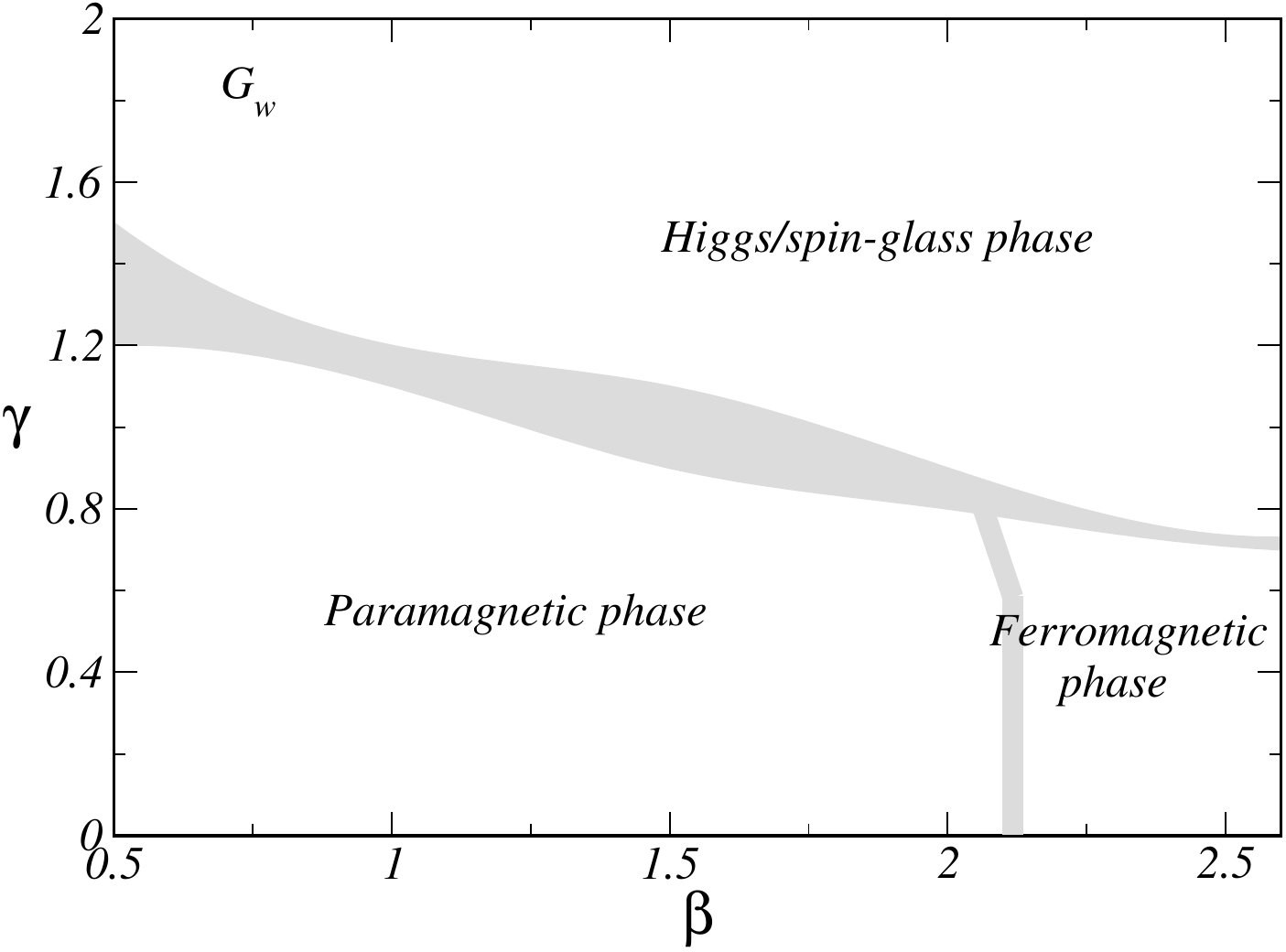}
\includegraphics[width=0.47\linewidth,clip]
{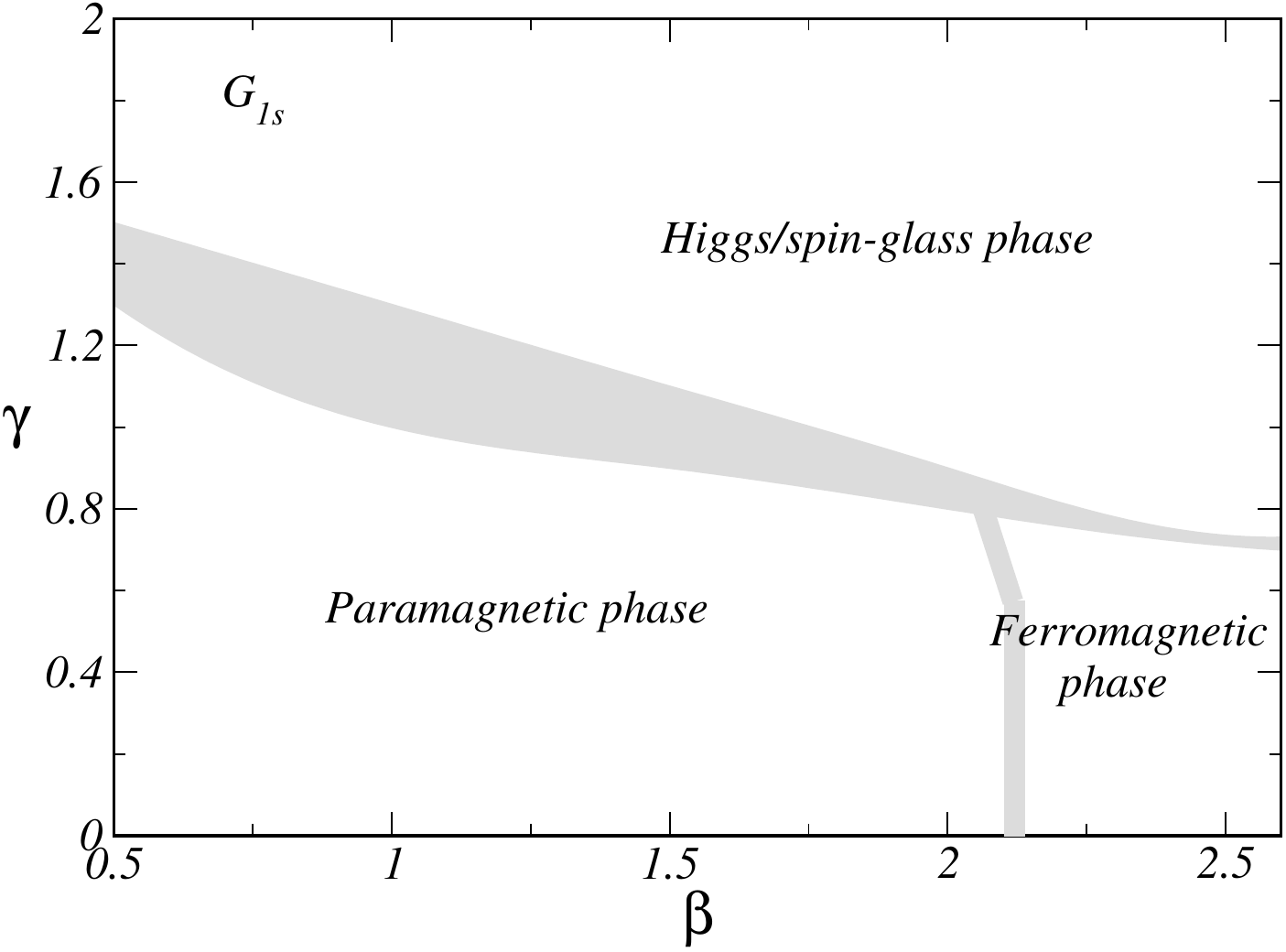}
\caption{Approximate phase diagram of the finite temperature $SU(2)$ gauge-Higgs model for $N_t=4$ obtained from the critical behavior of the overlap operator.}
\label{fig:phase diagram_overlap}
\end{figure}

\begin{itemize}
\item 
The two versions, $G_w$ and $G_{1s}$, of the overlap operator vanish in the confinement and deconfinement regions, whereas they take non-zero values in the Higgs phase. In this sense the overlap operator is as good order parameter at finite temperature as it is at zero temperature. In Fig.\ref{fig:phase diagram_overlap}
we show the resulting phase diagram of the $SU(2)$ gauge-Higgs LGT. We did not attempt to locate the critical lines with high precision as it requires so large $n_{\rm sym}$ and volumes, that are beyond our present computer capabilities. The borders of the shaded regions are lower and upper numerical limits of the critical lines. These regions, obtained respectively from $G_w$ and $G_{1s}$, are in a reasonable agreement with each other as one can see from two plots in Fig.\ref{fig:phase diagram_overlap}. 
\item 
Both temporal and spatial FM operators take on small but non-zero values in the confinement phase of the theory. In the Higgs phase of the theory both FM operators also tend to a constant which is substantially larger than in the confinement phase. In the deconfined phase both operators exhibit surprisingly similar behavior and tend to zero with increasing size $R$. However, large error bars do not allow us to make an unambiguous conclusion as to the vanishing 
of FM operators in the deconfined phase. 
\end{itemize}

Two further comments are in order. First, a detailed
study of the critical behavior is beyond the scope of the present work as it would require large spatial volumes and very big statistics. Nevertheless, 
it would be interesting and important to investigate the critical regions by computing the overlap susceptibility and by verifying the collapse of the FM operators in the spirit of Ref.~\cite{FM_86_MC}. This would give access to an universality class of the phase transitions.   

The second comment concerns the behavior of the spatial FM operator in the deconfined phase. As follows from our simple analytic calculations, we would rather expect that this operator stays constant in all phases of the theory including the deconfined one. To our surprise, we found that this operator behaves almost like the temporal one and might vanish in the deconfined region. If such a scenario were confirmed, it would imply that some important configurations are missing in our formulas for the expectation value of the Wilson line in the deconfined phase. Of course, one cannot exclude the possibility that both operators are very small but non-vanishing in the deconfined phase. 
We think it is worth checking these possible scenarios in the simpler $Z(2)$ gauge-Higgs model at finite temperature. By making use of larger lattices and collecting bigger statistics one could try to get   
unambiguous predictions as to the properties of the FM operators at finite temperature. This work is in progress.

{\bf Acknowledgments}. We thank M.P.~Forsstr\" om for useful information about 
the properties of the FM operator. The authors acknowledge support from INFN/NPQCD project.
This work is partially supported by ICSC -- Centro Nazionale di Ricerca in High Performance Computing, Big Data and Quantum Computing, funded by European Union -- NextGenerationEU. Most numerical simulations have been performed on the CSN4 cluster of the Scientific Computing Center at INFN-Pisa.
B.A. is grateful to Juan Jos\'e Alonso for useful advice on simulation methods for spin-glasses.

\end{document}